\documentclass[aps,prl, reprint, showpacs,showkeys,superscriptaddress,preprintnumbers,nofootinbib]{revtex4-1}
\usepackage{amsmath,amssymb,bm,mathrsfs}
\usepackage{hyperref}
\usepackage{url}
\usepackage{graphicx}
\usepackage{color}
\usepackage{mathtools}

\definecolor{grassgreen}{cmyk}{0.77,0,1,0.05}

\newcommand{\sml}[1]{\textcolor{black}{ #1}}

\newcommand{\dyc}[1]{\textcolor{black}{ #1}}

\newcommand{\yk}[1]{\textcolor{black}{ #1}}

\begin{document}

\preprint{\tt CERN-TH-2022-167}

\title{
Axion Quality Problem and Non-Minimal Gravitational Coupling \\
in the Palatini Formulation
}

\author{Dhong Yeon Cheong}
\email{dhongyeon@yonsei.ac.kr}
\affiliation{Department of Physics and IPAP, Yonsei University, Seoul 03722, Republic of Korea}

\author{Koichi Hamaguchi}
\email{hama@hep-th.phys.s.u-tokyo.ac.jp}
\affiliation{Department of Physics, University of Tokyo, 
Tokyo 113--0033,
Japan}
\affiliation{
Kavli Institute for the Physics and Mathematics of the Universe (Kavli
IPMU), University of Tokyo, Kashiwa 277--8583, Japan
}

\author{Yoshiki Kanazawa}
\email{kanazawa@hep-th.phys.s.u-tokyo.ac.jp}
\affiliation{Department of Physics, University of Tokyo, 
Tokyo 113--0033,
Japan}

\author{Sung Mook Lee}
\email{sungmook.lee@yonsei.ac.kr}
\affiliation{Department of Physics and IPAP, Yonsei University, Seoul 03722, Republic of Korea}
\affiliation{Theoretical Physics Department, CERN, 1211 Geneva 23, Switzerland}

\author{Natsumi Nagata}
\email{natsumi@hep-th.phys.s.u-tokyo.ac.jp}
\affiliation{Department of Physics, University of Tokyo, 
Tokyo 113--0033,
Japan}

\author{Seong Chan Park}
\email{sc.park@yonsei.ac.kr}
\affiliation{Department of Physics and IPAP, Yonsei University, Seoul 03722, Republic of Korea}
\affiliation{Korea Institute for Advanced Study, Seoul 02455, Republic of Korea}

\begin{abstract}

In axion models, the global U(1) Peccei-Quinn (PQ) symmetry is explicitly broken by non-perturbative effects of gravity, such as axionic wormholes. The gravitational violation of the PQ symmetry due to wormholes is large enough to invalidate the PQ mechanism, which is entitled as the axion quality problem. Recently, a novel solution to this quality problem was suggested, where the non-minimal coupling of the axion field to gravity $\xi$ is introduced to suppress the wormhole contribution. 
In this work, we revisit the problem in a different but equally valid formulation of gravity, namely the Palatini formulation, where the Ricci scalar is solely determined by connection. We first find the axionic wormhole solution in the Palatini formulation, taking the full dynamical radial mode as well as the axial mode, then show that the quality problem is still resolved with the non-minimal coupling $\xi$. The requested lower bound of $\xi$ in the Palatini formulation turns out to be slightly higher than that in the metric formulation.

\end{abstract}

\maketitle

%%%%%%%%%%%%%%%%%%%%%%%
\section{Introduction}
%%%%%%%%%%%%%%%%%%%%%%%

\dyc{The axion} \cite{Weinberg:1977ma,Wilczek:1977pj}, \yk{which is} a pseudo-Nambu-Goldstone boson associated with the spontaneous breaking of the global U(1) Peccei-Quinn (PQ) symmetry \cite{Peccei:1977hh,Peccei:1977ur}, is introduced as a solution to the strong CP problem. Below the QCD scale, the axion field obtains a periodic potential induced by QCD instantons to settle down to an exactly CP-conserving vacuum.
However, even a tiny violation of this symmetry can \dyc{jeopardize} the PQ solution to the strong CP problem, \dyc{leading to the mechanism being extremely sensitive to the quality of the PQ symmetry \cite{Dine:2022mjw}}.

\sml{From an effective field theory perspective, we expect that there should be higher dimensional operators of the form $ \Delta V \sim c\, \Phi^{n}/M_{P}^{n-4} \sim \frac{c  f_{a}^{n}}{M_{P}^{n-4}} \cos \left( \frac{n \phi}{f_{a}} + \delta \right)$ where $\Phi$ is a complex PQ scalar and $\phi$ is the associated axion field with a decay constant $f_{a}$ \cite{Hook:2018dlk}.} 
\sml{Indeed}, it is expected that gravity may spoil any global symmetry \sml{at \dyc{the} least}~\cite{Kallosh:1995hi,Banks:2010zn,Witten:2017hdv,Harlow:2018jwu,Harlow:2018tng} and the PQ symmetry may not be an exception. This % sml gravitational 
violation of the U(1) symmetry generally displaces the axion field from the CP-conserving minimum of the axion potential. If the deviation is too large, the induced value of the neutron electric dipole moment becomes incompatible with the experimental upper bound~\cite{Abel:2020pzs}, and hence the PQ mechanism does not work as a viable solution to the strong CP problem. This is called the `axion quality problem'~\cite{Kamionkowski:1992mf, Holman:1992us, Barr:1992qq}
\dyc{and there have been many possible solutions suggested in the literature~\cite{Witten:1984dg,Randall:1992ut,Cheng:2001ys,Izawa:2002qk,Choi:2003wr,Lillard:2018fdt}}.

Meanwhile, an axionic wormhole~\cite{Giddings:1987cg, Lee:1988ge, Kallosh:1995hi}, which is a gravitational instanton in Euclidean spacetime, is a well-known %example 
\sml{case} that explicitly shows the gravitational violation of the U(1) symmetry. It has a non-trivial topology characterized by a global U(1) charge and induces effective local operators that violate the U(1) symmetry \yk{through computing quantum transition amplitudes in a semi-classical approximation} \cite{Coleman:1988cy, Giddings:1988cx, Rey:1989mg, Abbott:1989jw}. (See also Refs.~\cite{Alonso:2017avz, Hebecker:2018ofv}.) \dyc{Assuming these non-perturbative effects as the  source of symmetry breaking, the coefficients of these operators have a characteristic exponentially suppressed %factor 
\sml{term consisting of a coefficient of order $c \sim e^{-S}$} with $S$ corresponding to the wormhole action \sml{\cite{Abbott:1989jw,Coleman:1989zu}}.}\footnote{\yk{See, for example, Ref. \cite{Svrcek:2006yi} for other sources of explicit PQ-violation such as stringy instantons.}}  
In order to \dyc{{issue}} the quality problem, the wormhole action should be large, $S \gtrsim 190$, 
for its effect to be suppressed enough \cite{Kallosh:1995hi, Alvey:2020nyh}.

Whether these wormholes induce the quality problem depends sensitively on the UV model that realizes the axion at low energies \cite{Alvey:2020nyh}. In a simple model with a periodic scalar field~\cite{Giddings:1987cg,Lee:1988ge}, the wormhole action is $S \sim \yk{n} M_{P}/f_{a}$; thus the value becomes large enough when the decay constant is small as $f_a \lesssim 10^{16}~\rm{GeV}$.
However, the result drastically changes if the axion corresponds to a phase component of a complex scalar field together with a dynamical radial component, $\Phi = f e^{i \theta}/\sqrt{2}$.
In this case, the field value of $f$ stays near the Planck scale at the throat of the wormhole, and the size of the wormhole also becomes close to the Planck length. 
 The wormhole action now  scales logarithmically as $ S \sim \yk{n} \log (M_{P}/f_{a})$ and it cannot grow sufficiently. Therefore, we still suffer from the quality problem \cite{Kallosh:1995hi,Alonso:2017avz, Alvey:2020nyh}.

A recent work \cite{Hamaguchi:2021mmt} suggested a \dyc{novel} solution to the quality problem even in the presence of the dynamical radial field, by introducing a non-minimal gravitational coupling of a complex scalar field, $\xi$. It is found that the quality problem is avoided for $\xi \gtrsim 2 \times 10^3$ having a sufficiently large wormhole action. We take this solution seriously, as the non-minimal coupling $\xi$ is allowed in any effective theory as long as the term is consistent with the symmetries of the theory. Indeed, the implications of the term have been widely explored in inflationary cosmology~\cite{Futamase:1987ua,Bezrukov:2007ep, Park:2008hz,Bauer:2008zj,Tamanini:2010uq,Hertzberg:2010dc,Hamada:2014iga, Hamada:2014wna, Jinno:2018jei, Jinno:2019und, Cheong:2021kyc, Cheong:2022gfc}.

One potential loophole of the suggested solution in Ref.~\cite{Hamaguchi:2021mmt} is noted: 
gravity can be equally valid when formulated in  different ways. Indeed,  distinctively from the conventional metric formulation taken in Ref.~\cite{Hamaguchi:2021mmt}, we can alternatively choose the Palatini formulation of gravity~\cite{Einstein:1925,Ferraris:1982} that takes the connection as an %quantity 
independent degree of freedom apart from the metric.
While both the metric and Palatini formulations 
are equivalent within the so-called minimal gravity model of the Einstein-Hilbert action, the equivalence is generally broken in non-minimal models~\cite{Ferraris:1992dx,Magnano:1993bd}. 
Since we are not able to \dyc{distinctively rule out either formulation} at the current stage of our experimental knowledge, in this paper, we pursue to reconsider the Palatini formulation and examine if the noble solution to the quality problem remains valid. \footnote{{The intrinsic difficulty of the experimental probes becomes more transparent when one considers the Einstein frame. In this frame, the gravity sector becomes canonical in both formulations of gravity and all effects are recast to the scalar potential deformation, which non-minimally couples to the Ricci scalar in the Jordan frame. As long as we are not able to probe scalar field values \dyc{large enough for} the deformation of the potential \dyc{to be} significant, $f \gtrsim M_{P} / \xi $, the differences in observables become negligible. On the other hand, for $f(R)$ gravity, there could be constraints for Palatini formulation coming from the fact that it violates the equivalence principle \cite{Olmo:2006zu,Wu:2018idg}.}}

%%%%%%%%%%%%%%%%%%%%%%%
\section{Model}
%%%%%%%%%%%%%%%%%%%%%%%

The action for a complex scalar field $\Phi = \frac{f}{\sqrt{2}}e^{i \theta}$ with non-minimal coupling with gravity in the Palatini formulation is given as
\begin{align}
    S
    &= \int d^4x \sqrt{|g|} \left[
    - \frac{M^2+2\xi|\Phi|^2}{2}R(\Gamma) +|\partial_{\mu}\Phi|^2+V(|\Phi|) \right], \label{eq:action_def}
\end{align}
where $g=\det g_{\mu\nu}$, the mass parameter $M$ is defined as $M^2 = M_P^2 - \xi f_a^2$ with the Planck mass $M_P = 1/\sqrt{8 \pi G} \approx 2.4\times 10^{18}~{\rm GeV}$ and $V(|\Phi|)=\lambda \left(|\Phi|^2 -f_a^2/2 \right)^2$. The vacuum expectation value of $\Phi$ is $\sqrt{\langle \Phi^\dagger \Phi \rangle} =f_a/\sqrt{2}$ such that the gravitational coupling becomes canonical at the vacuum. In order to keep the correct sign of the kinetic term of the graviton, we request $M^2 \geq 0$, or equivalently $\xi \leq M_P^2/f_a^2$. The Ricci scalar $R(\Gamma)$ is obtained from the Ricci tensor, which is explicitly given as 
\begin{align}
    R_{\mu \nu}(\Gamma
    ) = \partial_{\mu} \Gamma^{\lambda}_{\lambda \nu} - \partial_{\lambda} \Gamma^{\lambda}_{\mu \nu} + \Gamma^{\lambda}_{\mu \sigma} \Gamma^{\sigma}_{\lambda \nu} - \Gamma^{\sigma}_{\mu \nu} \Gamma^{\lambda}_{\lambda \sigma}.
\end{align}
Due to the absence of second-order derivatives, unlike the metric formulation, the Gibbons-Hawking-York boundary term \cite{York:1972sj, Gibbons:1976ue} is not necessary for the Palatini formulation. 
We take the Euclidean geometry with spherical symmetry, $ds^2 = dr^2 + a(r)^2 d^2\Omega_3$, where $r$ is the Euclidean time, $d^2\Omega_3$ is the line element on the three-dimensional unit sphere and $a$ is the radius of the sphere. We assume that $f$ and $\theta$ depend only on $r$, taking the spherical symmetry into account.

%%%%%%%%%%%%%%%%%%%%%%%
\section{Analysis in Palatini formulation}
%%%%%%%%%%%%%%%%%%%%%%%

%%%%%%%%%%%%%%%%%%%%%%%
\subsection{Wormhole Solutions}
%%%%%%%%%%%%%%%%%%%%%%%

In a semi-classical approximation, a wormhole is a saddle-point solution in the Euclidean path integral, with a boundary condition on the canonical momentum of the axion field \sml{\cite{Kallosh:1995hi,Alonso:2017avz,Hebecker:2018ofv}}. Expanding Eq.~\eqref{eq:action_def}, we have
\begin{align}
	S = \int d^{4}x
	\sqrt{|g|} & \left[  
	- \frac{M_{P}^{2}}{2} \Omega^{2}(f) R + \frac{1}{2} (\partial_{\mu} f)^{2} \right. \nonumber \\
	& \left.
     \quad\quad\quad \quad+\frac{1}{2} f^{2} (\partial_{\mu} \theta )^{2}  + V(f) \right] ~, \label{eq:action}
\end{align}
where
\begin{align}
    \Omega^2 (f) \equiv 1 + \frac{\xi ( f^2 - f_a^2 )}{M_P^2}.
\end{align}
The variation with respect to $\theta$ gives 
\begin{align}
 \partial_{\mu} \left( \sqrt{g} f^{2} \partial^{\mu}\theta \right)= 0,
\end{align}
motivating us to define a conserved current,
\begin{align}
	J^{\mu} = \sqrt{g} f^{2} \partial^{\mu}\theta(r),
\end{align}
which is associated with shift symmetry. The conserved charge is quantized as
\begin{align}
     2\pi^2 a^3 f^2 \theta^{\prime}(r) = n \in \mathbb{Z}, \label{eq:constraint}
\end{align}
due to the $2\pi$-periodicity of the axion and the spherical symmetry of the metric. The prime($\prime$) denotes the derivative with respect to $r$. The integer $n \neq 0$ corresponds to the charge going through the wormhole and characterizes a wormhole solution.

To take this constraint into account properly, one can either impose the condition Eq.~\eqref{eq:constraint} from the beginning and plug this into the action in Eq.~\eqref{eq:action}, or introduce the constraint as a Lagrange multiplier \cite{Lee:1988ge,Alvey:2020nyh}.\footnote{For rigorous treatments on this point, see Ref.~\cite{Coleman:1989zu}.} For sure, both ways give identical results. Note that in the former way, the kinetic term of $ \theta $ gives an additional effective potential that corresponds to a `centrifugal force'. As we will see, this is the origin of the large field value at the wormhole throat once the radial field $f$ is taken to be dynamical.

From the variational principle with respect to $\Gamma^{\lambda}_{\mu\nu}$, we find
\begin{align}
    \nabla_{\lambda} [ M_P^2 \Omega^2(f) \sqrt{g} g^{\mu \nu} ] = 0,
    \label{eq:affine}
\end{align}
or
\begin{align}
    &\Gamma^{\lambda}_{\mu \nu} 
    = \bar{\Gamma}^{\lambda}_{\mu \nu} 
    + \delta^{\lambda}_{\mu} \partial_{\nu} \omega %(f)
    + \delta^{\lambda}_{\nu} \partial_{\mu} \omega %(f)
    - g_{\mu \nu} \partial^{\lambda} \omega %(f)
    ,
    \label{eq:affine2}
\end{align}
where
\begin{align}
    \omega (f) \equiv \log \vert \Omega (f) \vert.
\end{align}
The first term in Eq.~\eqref{eq:affine2} is the Levi-Civita connection,
\begin{align}
    \bar{\Gamma}^{\lambda}_{\mu \nu} 
    = \frac{1}{2}g^{\lambda \alpha}(
    g_{\mu \alpha, \nu} + g_{\alpha \nu, \mu} - g_{\mu \nu, \alpha}
    )~,
\end{align}
while the last three terms are additional terms that are absent in the metric formulation and depend on the non-minimal coupling. We also obtain
\begin{align}
     R & = g^{\mu \nu} R_{\mu \nu} (\Gamma%, \partial \Gamma
    )     \nonumber
    \\
    & = -6 \left(
    \frac{a^{\prime \prime}}{a} + \frac{a^{\prime 2}}{a^2} - \frac{1}{a^2}
    \right)
    - 6 \left[
    {\omega}^{\prime 2} + {\omega}^{\prime\prime}+ 3 \frac{ a^{\prime}}{a} {\omega}^{\prime}
    \right].
\end{align}
From the variation with respect to $g_{\mu \nu}$, we find
\begin{align}
    &\Omega^2 \left[ a^{\prime 2} - 1 +
    2 a a^{\prime} {\omega}^{\prime} +a^2 {\omega}^{\prime 2}
    \right]
    \nonumber
    \\
    & \quad = -\frac{a^2}{3 M_P^2} \left[
    -\frac{1}{2} f^{\prime 2} + V(f) + \frac{n^2}{8 \pi^4 f^2 a^6}
    \right],     \label{eq:metric-1} 
\end{align}
and
\begin{align}
    &\Omega^2 \left[ 2 a a^{\prime \prime} + a^{\prime 2} -1  
    + 4 a a^{\prime} {\omega}^{\prime} + a^2 {\omega}^{\prime 2}
    + 2 a^2 {\omega}^{\prime\prime} \right] 
    \nonumber\\
    &\quad = -\frac{a^2}{M_P^2} \left[
    \frac{1}{2} f^{\prime 2} + V(f) - \frac{n^2}{8 \pi^4 f^2 a^6}
    \right],
    \label{eq:metric-2}
\end{align}
while from the variation with respect to $f$, we obtain
\begin{align}
    &f^{\prime \prime} + 3 \frac{a^{\prime}}{a} f^{\prime}  - \frac{dV}{df} + \frac{n^2}{4 \pi^4 f^3 a^6} \nonumber  \\
    &\quad = 6 \xi f \left[
    \frac{a^{\prime \prime}}{a} + \frac{a^{\prime 2}}{a^2} -\frac{1}{a^2}
    + 
    {\omega}^{\prime 2} + {\omega}^{\prime\prime} + 3 \frac{a^{\prime}}{a} {\omega}^{\prime}
    \right].
    \label{eq:f}
\end{align}

We solve Eqs. \eqref{eq:metric-1}, \eqref{eq:metric-2} and \eqref{eq:f} numerically with the following boundary conditions,
\begin{align}
    a^{\prime} (0) = 0, &&  f^{\prime} (0) = 0, && f (\infty) = f_a.
\end{align}
For convenience, we introduce dimensionless parameters,
\begin{align}
    \rho \equiv \sqrt{3 \lambda} M_P r, && A \equiv \sqrt{3 \lambda} M_P a, && F \equiv \frac{f}{\sqrt{3} M_P}.
\end{align}

\begin{figure} [t]
    \centering
    \includegraphics[clip, width = 0.45\textwidth]{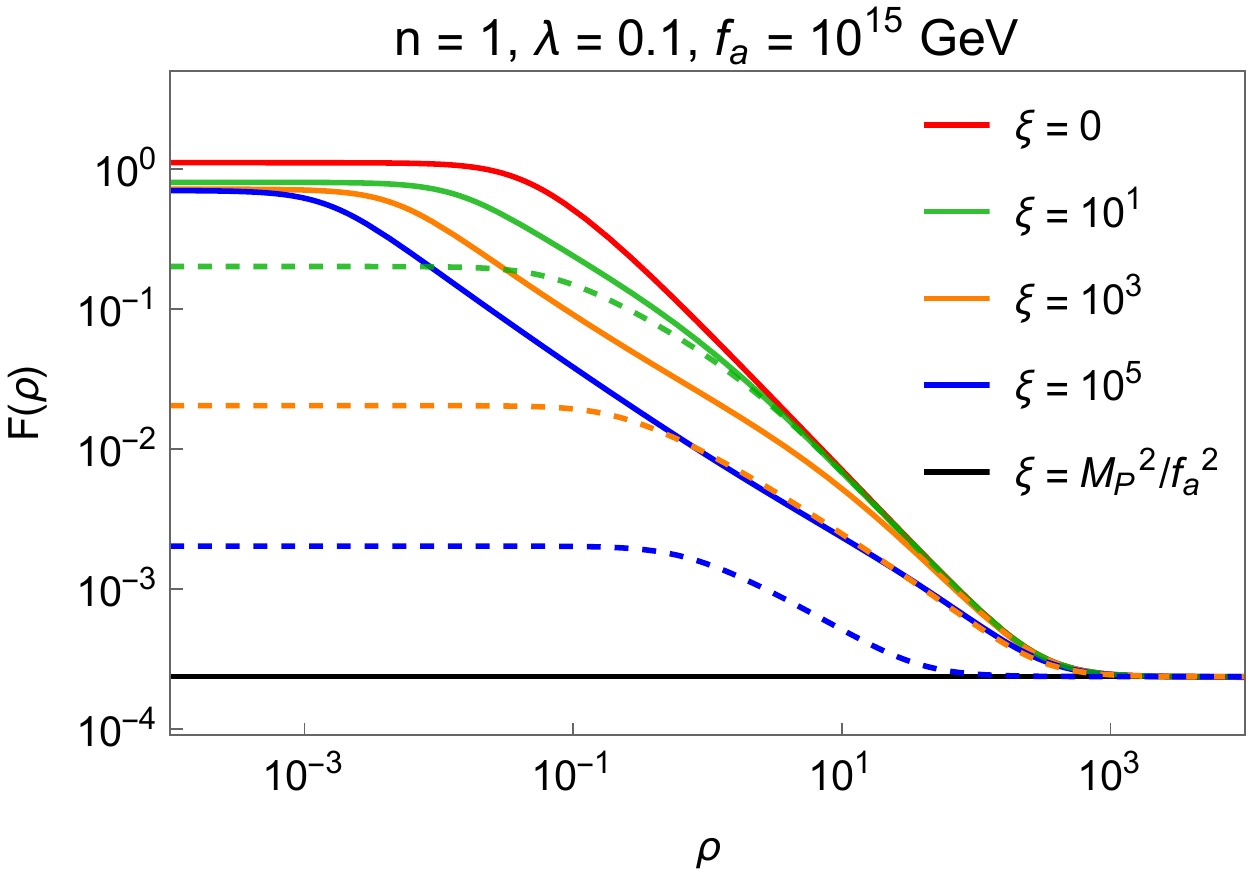}
    \caption{$F(\rho)$ for several values of $\xi$. The solid and dashed lines are for the Palatini and metric formulations, respectively. 
    }
    \label{fig:F_rho}
\end{figure}

\begin{figure} [t]
    \centering
    \includegraphics[clip, width = 0.45\textwidth]{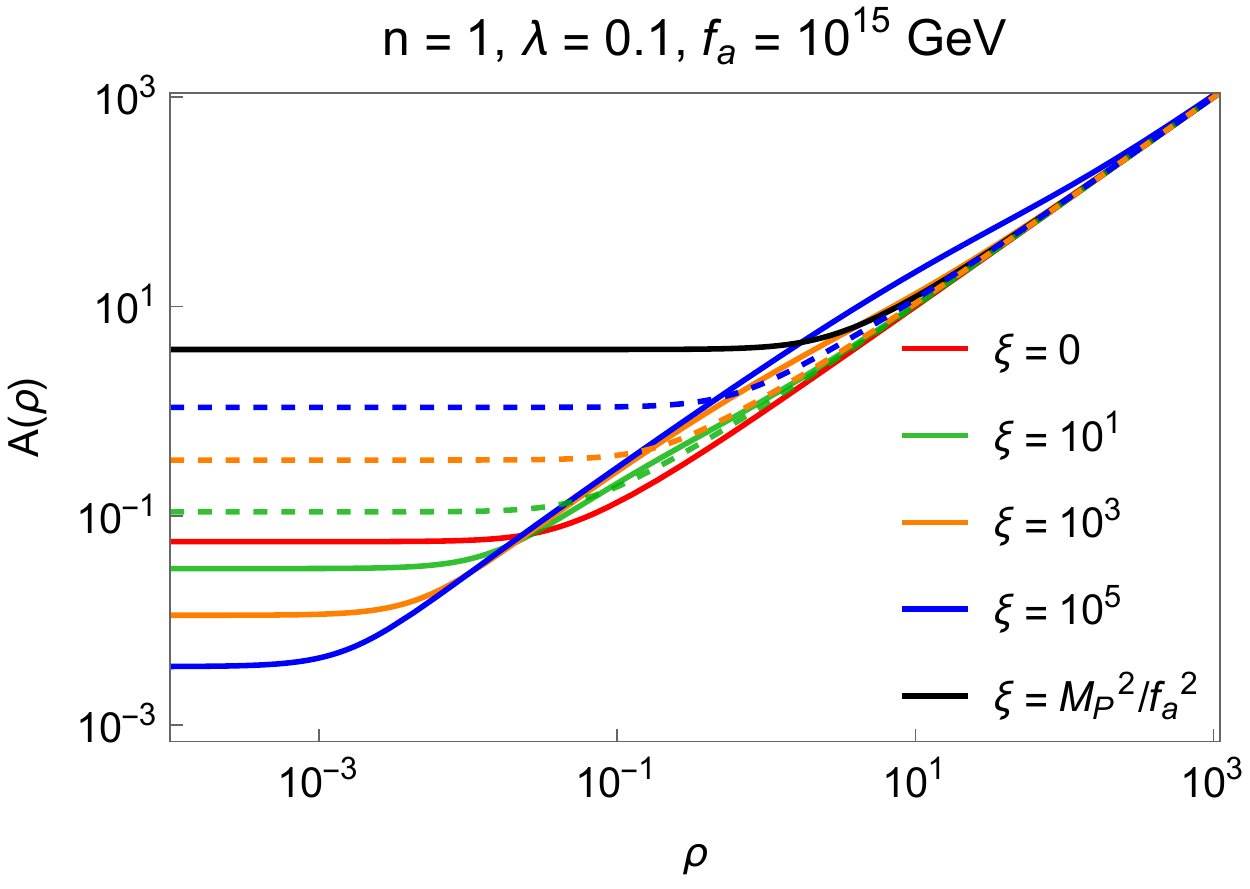}
    \caption{$A(\rho)$ for several values of $\xi$. The solid and dashed lines are for the Palatini and metric formulations, respectively.
    }
    \label{fig:A_rho}
\end{figure}

We show the graphs of $F(\rho)$ and $A(\rho)$ in Fig.~\ref{fig:F_rho} and Fig.~\ref{fig:A_rho}, respectively. First, for $\xi = 0$ and $\xi = M_P^2 / f_a^2$, both the metric (dashed) and Palatini (solid) formulations give \dyc{identical} results. In particular, as found in Ref.~\cite{Hamaguchi:2021mmt}, the solution for $\xi = M_P^2 / f_a^2$ (the induced gravity model) is identical to the Giddings-Strominger (GS) wormhole corresponding to $f(r) = f_a$ \cite{Giddings:1987cg}.

%%%%%%%%%%%%%%%%%%%%%%%%%%%%%%%%%%%%%%%%%%%%%%%%%%%%
%%%%%%%%%%%%%%%%%%%%%%%%%%%%%%%%%%%%%%%%%%%%%%%%%%%%

%%%%%%%%%%%%%%%%%%%%%%%%%%%%%%%%%%%%%%%%%%%%%%%%%%%%
%%%%%%%%%%%%%%%%%%%%%%%%%%%%%%%%%%%%%%%%%%%%%%%%%%%%

To understand this trend, it is illuminating to consider the action Eq.~(\ref{eq:action}) in the Einstein frame, with a metric redefinition $ g_{\mu\nu} \rightarrow \Omega^{-2} g_{\mu\nu} $ \cite{Accetta:1989gh}. In this frame,
\begin{align}
	S_{E} = \int d^{4}x \sqrt{g} & \left[  - \frac{M_{P}^{2}}{2} \left(R - \frac{ 3 \zeta (\partial_{\mu} \Omega^{2})^{2} }{ 2 \Omega^{4}}  \right)   \right. \nonumber
	\\ & ~~ \left.
	+ \frac{1}{2 \Omega^{2}} (\partial_{\mu}f)^{2} + 
	\frac{1}{2 \Omega^{2}} f^{2} (\partial_{\mu} \theta )^{2}  + \frac{V}{\Omega^{4}} \right] ~,
\end{align}
where $\zeta = 0 ,1$ in the Palatini and metric formalisms, respectively, and we have not canonicalized the kinetic term of $f$ as it is irrelevant for the discussion. Note that, in the induced-gravity limit with $\Omega^{2} =  f^{2} / f_{a}^{2} $, the coefficient of $(\partial_{\mu} \theta)^{2}$ becomes constant. This shows the decoupling between $f$ and $\theta$ in this limit, hence the reason for staying in the GS solution with the absence of the additional force more explicitly.

\begin{figure} [t]
    \centering
    \includegraphics[clip, width = 0.45\textwidth]{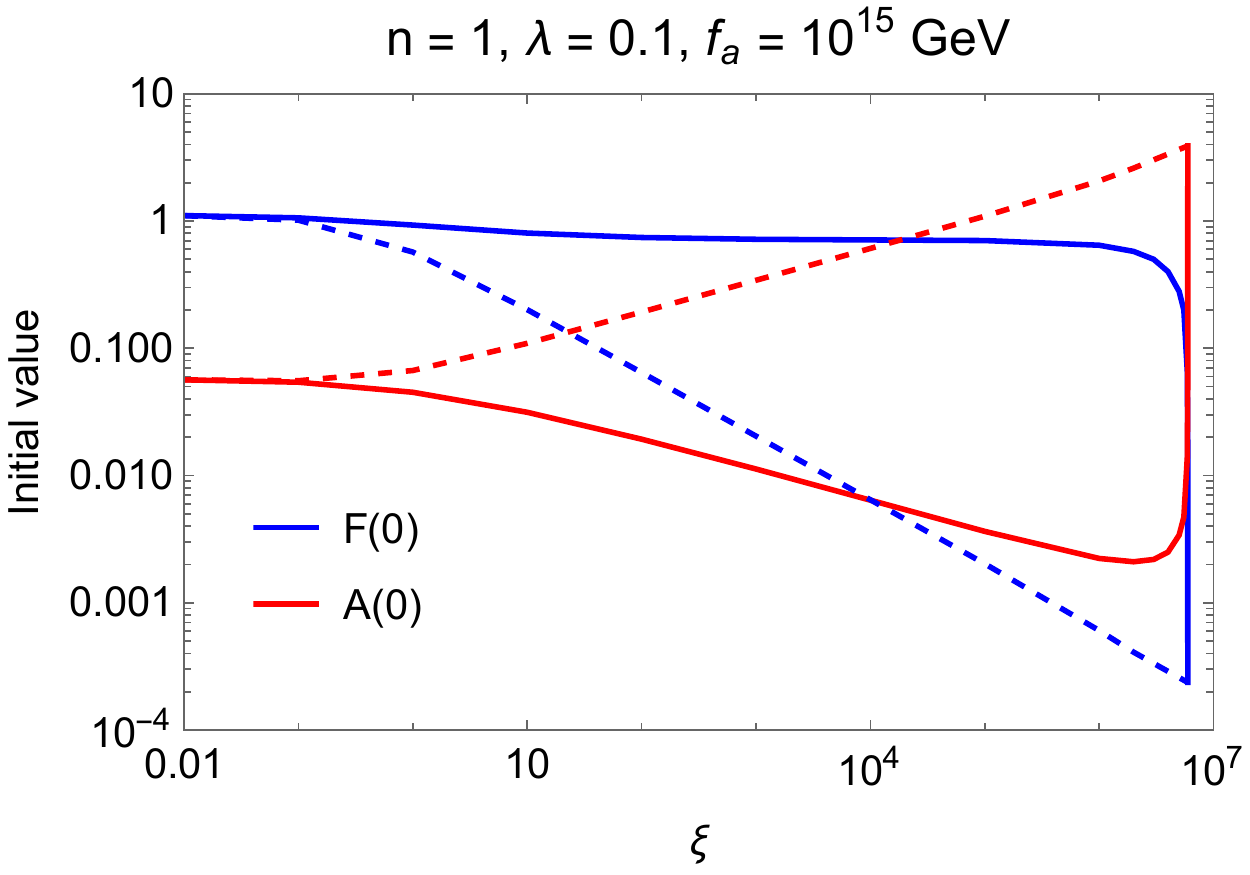}
    \caption{$F(0)$ and $A(0)$ as functions of $\xi$. The solid and dashed lines are for the Palatini and metric formulations, respectively. 
    }
    \label{fig:F0_and_A0}
\end{figure}

For intermediate values of $\xi$, the behavior of the solutions differs between the metric and Palatini formulations. The difference is especially \dyc{noticeable} at the wormhole throat, as shown in Fig.~\ref{fig:F0_and_A0}. This figure shows that in the Palatini case, the radial field $f$ stays near the Planck scale (corresponding to $F(0) \sim 1$) \dyc{except for extremely close values} to $\xi = M_P^2 / f_a^2$ and quickly transits to the GS wormhole near the induced-gravity limit $\xi = M_P^2 / f_a^2$. \dyc{As $\xi$ increases from 0, the radius of the wormhole throat $a(0)$ initially decreases and then increases drastically near $\xi = M_P^2 / f_a^2$ in the Palatini case, while it monotonically increases in the metric case.} \dyc{We have also confirmed that the $\lambda$-dependence of the wormhole solutions for small $ \lambda $s is negligible, resulting in identical wormhole geometries for both the metric and Palatini cases, leading to robust predictions.}

One may wonder if there is \dyc{no problem with the size of the wormhole throat being comparable to or even smaller than the Planck length.} In fact, such a large non-minimal coupling also introduces a perturbative unitarity cutoff $\Lambda_{J}$ to the theory, where $J$ denotes the Jordan frame \cite{Burgess:2009ea,Barbon:2009ya,Burgess:2010zq,Hertzberg:2010dc,Bezrukov:2010jz,Bauer:2010jg,Shaposhnikov:2020fdv,Antoniadis:2021axu}.\footnote{The cutoff scales in the Einstein frame, $\Lambda_{E}$, and in the Jordan frame, $\Lambda_{J}$, are related through a conformal factor $\Omega$, as $\Lambda_{J} = \Omega \Lambda_{E}$.} It is known that,\footnote{We note that the perturbative unitarity cutoff of the complex U(1) scalar is higher than the SU(2) scalar doublet as the SM Higgs. This difference is nicely discussed and summarized in the addendum of Ref.~\cite{Antoniadis:2021axu}.} for small field values
\begin{align}
    \Lambda_{J} \left( f \ll \frac{M_{P}}{\sqrt{\xi}} \right) \simeq \begin{dcases}
    M_{P} / \xi & \text{(metric)} \\
    M_{P} / \sqrt{\xi} & \text{(Palatini)}
    \end{dcases}
    \label{eq:cutoff}
\end{align}
while for large field values
\begin{align}
    \Lambda_{J} \left( f \gg \frac{M_{P}}{\sqrt{\xi}} \right) \simeq \begin{dcases}
    \xi f^{2}/M_{P} & \text{(metric)} \\
     \sqrt{\xi} f^{2}/M_{P} & \text{(Palatini)}.
    \end{dcases}
\end{align}
Near the throat, the field value of $f$ can be as large as the Planck scale, but the cutoff scale also increases as $\Lambda_{J}^{(M)} \sim \xi M_{P}$ and $\Lambda_{J}^{(P)} \sim \sqrt{\xi} M_{P}$ for the metric and Palatini cases, respectively, giving a cutoff much larger than the Planck scale. This partly justifies the self-consistency of our calculation\footnote{We, however, note in passing that although our calculation for wormholes is self-consistent, if we consider $f$ as an  inflaton, a large value of $\xi$ as in Eq.~\eqref{eq:xiinf} may still give rise to a unitarity problem at the stage of preheating for the metric formulation~\cite{Ema:2016dny, Sfakianakis:2018lzf, Ema:2021xhq}. On the other hand, if the quartic coupling $\lambda$ is less than $\mathcal{O}(1)$, the unitarity problem may not occur for the Palatini formulation, for any value of $\xi$~\cite{Ema:2021xhq}. This difference between the two formulations is highly related to the dependence of the cutoff scale on $\xi$ as given in Eq.~\eqref{eq:cutoff}.
} in the semi-classical regime while the wormhole throat becomes smaller than the Planck length for large $\xi$ in the Palatini case, as depicted in Fig.~\ref{fig:A_rho}.

%%%%%%%%%%%%%%%%%%%%%%%%%%%%%%%%%%%%%%%%%%%%%%%%%%%%
%%%%%%%%%%%%%%%%%%%%%%%%%%%%%%%%%%%%%%%%%%%%%%%%%%%%

%%%%%%%%%%%%%%%%%%%%%%%
\subsection{Quality Problem}
%%%%%%%%%%%%%%%%%%%%%%%

%%%%%%%%%%%%%%%%%%%%%%%%%%%%%%%%%%%%%%%%%%%%%%%%%%%%
\begin{figure}[t]
    \centering
    \includegraphics[clip, width = 0.45 \textwidth]{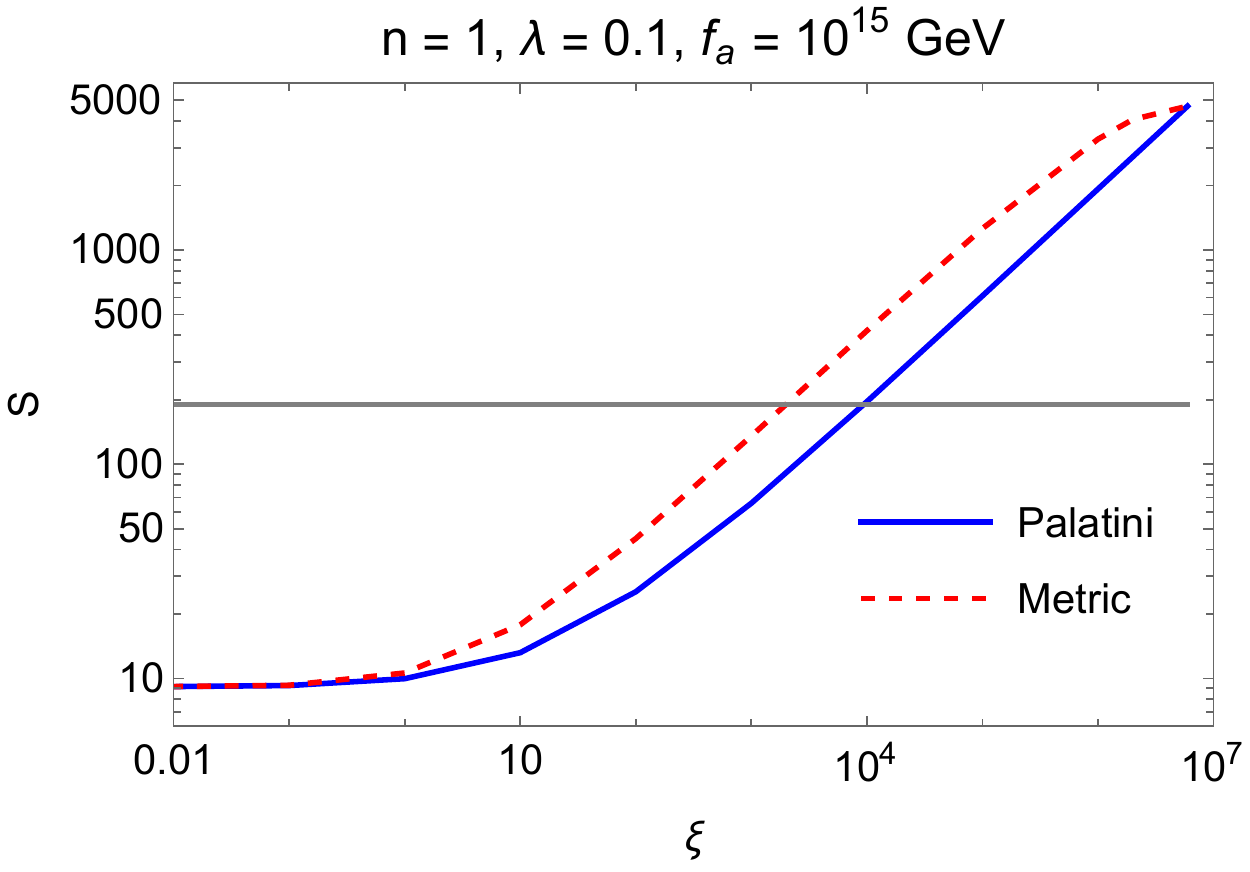}
    \caption{
    Values of wormhole action as a function of $\xi$. The blue-solid and red-dashed lines are for the Palatini and metric formulations, respectively. The horizontal gray line shows a lower bound to solve the quality problem, $S = 190$.
    }
    \label{fig:wormhole_action}
\end{figure}
%%%%%%%%%%%%%%%%%%%%%%%%%%%%%%%%%%%%%%%%%%%%%%%%%%%%

\yk{The PQ-violating operators are exponentially suppressed by the wormhole action. The wormhole action is almost proportional to the PQ charge $n$ for a given $\xi$. Thus, we focus on the contributions of the wormholes with $n=1$ to the quality problem.} 
We now compute the wormhole action,
\begin{align}
    S 
     = 2\pi^2 \int_0^{\infty} dr
    ~ a^{3} \left[
    f^{\prime 2} + 3 M_P^2 \Omega^2
    \left\{
    \frac{a^{\prime \prime}}{a} +  \frac{a^{\prime}}{a} {\omega}^{\prime}
    + {\omega}^{\prime\prime}
    \right\}
    \right].
\end{align}
In Fig.~\ref{fig:wormhole_action}, we show the wormhole action as a function of $\xi$ in both Palatini and metric formulations. For $0 < \xi < M_P^2/f_a^2$, the wormhole action in the Palatini formulation is found to be smaller than that in the metric formulation. For $\xi = 0$ and $\xi = M_P^2/f_a^2$, both formulations give the same result, as expected. Note that the quality problem can be solved for $S \gtrsim 190$ \cite{Kallosh:1995hi, Alvey:2020nyh}. This corresponds to $\xi \gtrsim 1 \times 10^4$ in the Palatini case, which is roughly an order of magnitude more stringent than that in the metric case, $\xi \gtrsim 2 \times 10^3$~\cite{Hamaguchi:2021mmt}. It is also worth noting that, for the Palatini case, the wormhole action increases even when the radius of the throat of the wormhole decreases.

%%%%%%%%%%%%%%%%%%%%%%%%%%%%%%%%%%%%%%%%%%%%%%%%%%%%
\begin{figure}
    \centering
    \includegraphics[clip, width = 0.45 \textwidth]{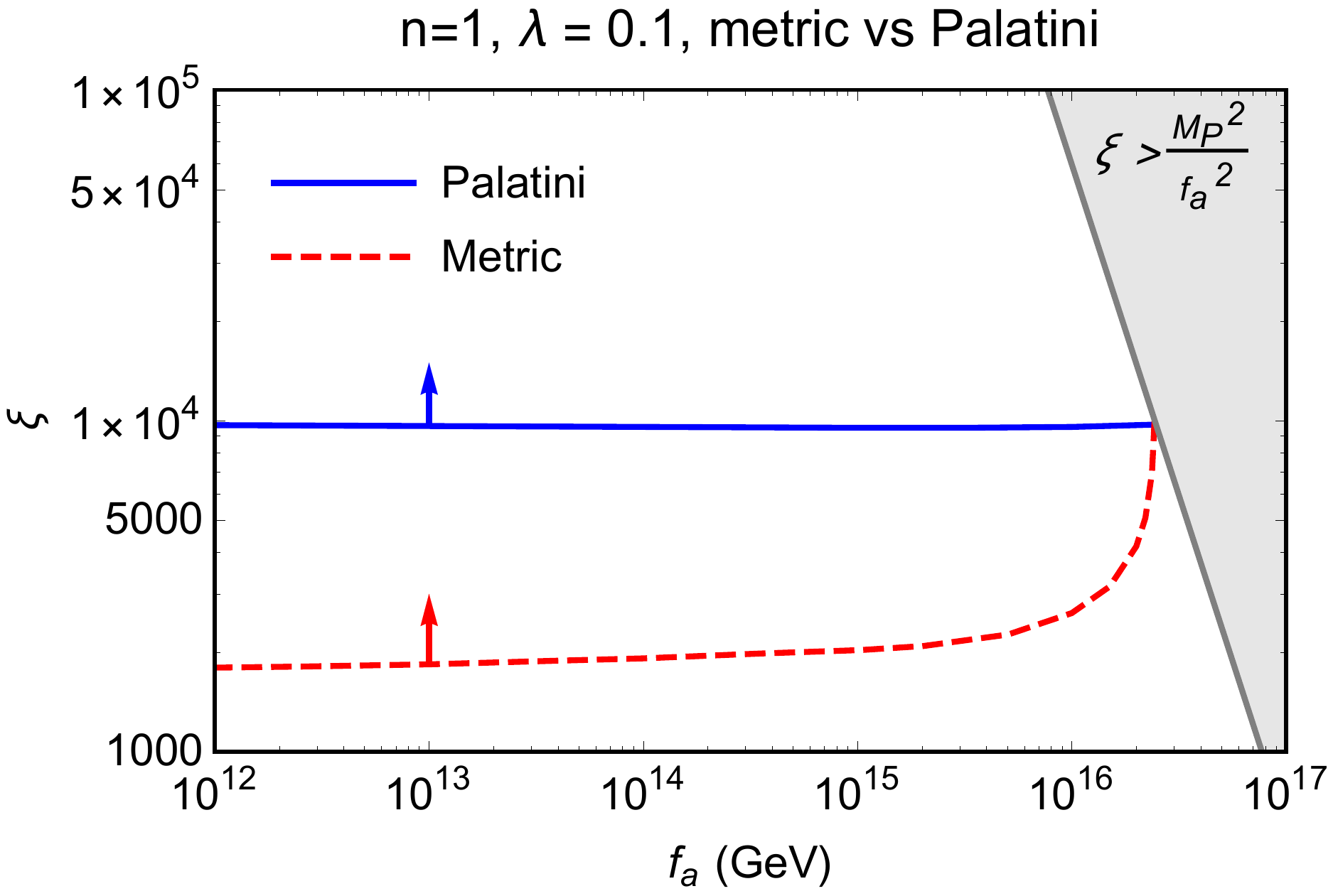}
    \caption{
    $f_a$-$\xi$ boundaries that marginally solve the quality problem in the Palatini (blue-solid) and metric (red-dashed) formulations. Regions above each line correspond to values that are safe from wormhole solutions spoiling the axion quality. The right gray region corresponds to values $\xi > M_P^2 / f_a^2 $. 
    }
    \label{fig:contour_diagram}
\end{figure}
%%%%%%%%%%%%%%%%%%%%%%%%%%%%%%%%%%%%%%%%%%%%%%%%%%%%

\begin{figure} [t]
    \centering
    \includegraphics[clip, width = 0.45\textwidth]{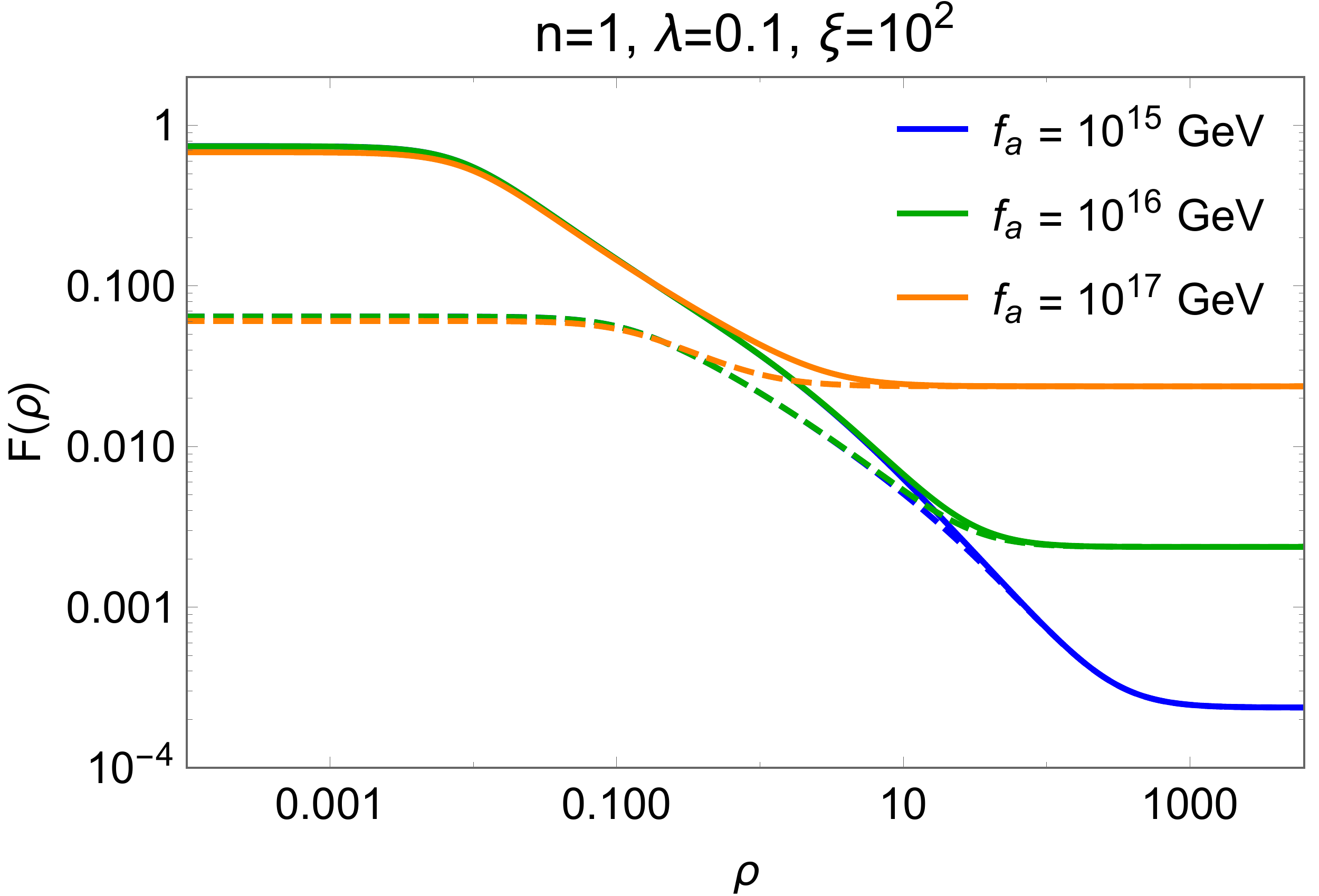}
    \caption{$F(\rho)$ for several values of $f_a$. The solid and dashed lines are for the Palatini and metric formulations, respectively.
  %  \TODO{It is better to add a line for $\xi = M_P^2/f_a^2$?}
    }
    \label{fig:F_rho_diff_fa}
\end{figure}

In Fig.~\ref{fig:contour_diagram}, we also depict the $f_a$-$\xi$ plane which shows the parameter space that solves the axion quality problem with axionic wormholes, for both Palatini (blue-solid) and metric (red-dashed) formulations. Note that the required values of $\xi$ not to have the quality problem hardly depend on the PQ symmetry breaking scale $f_a$. This $f_{a}$ independence may be understood as follows. As shown in Fig.~\ref{fig:F_rho_diff_fa}, for a fixed value of $\xi$, the solutions of $F(\rho)$ and $A(\rho)$ are almost independent of $f_{a}$ except for a large value of $\rho$. On the other hand, the contribution to the action $S$ is dominated by the region of small $\rho %\lesssim 1
$ near the wormhole throat. Thus, the region of large $\rho$ (which is sensitive to $f_{a}$) 
hardly contributes
to the $S$.

%%%%%%%%%%%%%%%%%%%%%%%%%%%%%%%%%%%%%%%%%%%%%%%%%%%%
%%%%%%%%%%%%%%%%%%%%%%%%%%%%%%%%%%%%%%%%%%%%%%%%%%%%

%%%%%%%%%%%%%%%%%%%%%%%%%%%%%%%%%%%%%
\section{Conclusion and discussions}
%%%%%%%%%%%%%%%%%%%%%%%%%%%%%%%%%%%%%

In this work, we have discussed the effect of the non-minimal coupling to gravity $\xi$ on the axion quality problem in the Palatini formulation. In this formulation, the affine connection and the metric are independent a priori. Hence, for non-zero $\xi$, the affine connection is different from the Levi-Civita connection of metric formulation, and so are the physical consequences compared to the usual metric formulation.

We have found that the presence of additional terms in the affine connection does affect the wormhole solution, resulting in a smaller wormhole action compared to that in the metric formulation. As a result, a larger value of the non-minimal coupling, $\xi \gtrsim 1 \times 10^4$, is required in order to avoid the quality problem. We have also checked \dyc{that} our calculation does not violate the perturbative unitarity of the theory with large non-minimal coupling thanks to large radial field values at the wormhole throat.

More specifically, as $\xi$ increases, the wormhole throat decreases at first while the wormhole action value increases. For larger $\xi$ near the induced gravity limit, there is a rapid convergence to the Gidding-Stronminger solution. This coincidence happens in both the metric and Palatini formulations, and could be understood as a decoupling between the axion $\theta$ and the radial mode $f$, as explicitly shown in the Einstein frame.

Lastly, as mentioned in the introduction, a scalar field with a large non-minimal coupling is sometimes considered \dyc{in the context of inflation which gives a consistent fit to CMB observations.} In our case, the radial field $f$ can play the role of the inflaton. To be consistent with the measurement of the scalar amplitude of the power spectrum $A_{s} \simeq 2.1 \times 10^{-9}$ \cite{Planck:2018jri} with about \dyc{60 e-folds}, we request a \dyc{substantial} value of $\xi$ for successful inflation~\cite{Bezrukov:2007ep, Bauer:2008zj},
\begin{align}
    \xi \simeq \begin{dcases}
    4.9 \times 10^{4} \sqrt{\lambda} & (\text{metric}) \\
    1.4 \times 10^{10} \lambda & (\text{Palatini}).
    \end{dcases}
    \label{eq:xiinf}
\end{align}
Note that a larger value of $\xi$ is needed for the Palatini case and it can easily satisfy the condition to solve the axion quality problem when $\lambda > 10^{-6}$.

%%%%%%%%%%%%%%%%%% Acknowledgements %%%%%%%%%%%%%%%%%%%%%%%%%%%%%%%%%%%%%%%
\section*{Acknowledgments}
NN would like to thank KIAS for inviting him to an online seminar, where this work was initiated. We thank Miguel Escudero, Fabrizio Rompineve, and Ryusuke Jinno for helpful discussions. This work is supported in part by the Grant-in-Aid for Innovative Areas (No.19H05810 [KH], No.19H05802 [KH], No.18H05542 [NN]), Scientific Research B (No.20H01897 [KH and NN]), Young Scientists (No.21K13916 [NN]), and JSPS Fellows (No.21J20445 [YK]). This work is supported in part by National Research Foundation grants funded by the Korean government (MSIT) (NRF-2019R1A2C1089334), (NRF-2021R1A4A2001897) [SCP] and Korea-CERN Theoretical Physics Collaboration and Developing Young High-Energy Theorists fellowship program (NRF-2012K1A3A2A0105178151) [SML]. SML is supported by the Hyundai Motor Chung Mong-Koo Foundation Scholarship.

%%%%%%%%%%%%%%%%%%%%%%%%%%%%%%%%%%%%%%%%%%%%%%%%%%%%%%%%%%%%%%%%%%%%%%%%%%%

%%%%%%%%%%%%%%%%% Ref %%%%%%%%%%%%%%%%%%%%%%%%
\bibliography{ref}
%%%%%%%%%%%%%%%%%%%%%%%%%%%%%%%%%%%%%%%%%%%%%%

\end{document}